\documentclass[12pt, notitlepage]{article}
\usepackage[utf8]{inputenc}
\usepackage[T1]{fontenc}
\usepackage{tgtermes}
\usepackage{amsfonts, amsmath, amsthm, amssymb}
\usepackage[top=1in,bottom=1in,left=1in,right=1in]{geometry}
\DeclareMathAlphabet{\mathpzc}{OT1}{pzc}{m}{it}
\usepackage[fleqn]{mathtools}
\usepackage{graphicx}
\usepackage{float}
\usepackage{authblk}
\usepackage{natbib}
\usepackage{enumitem}
\usepackage{multicol}
\usepackage{booktabs}
\usepackage{array}
\newcolumntype{P}[1]{>{\centering\arraybackslash}p{#1}}
\usepackage{threeparttable}
\usepackage{multirow}
\usepackage{setspace}
\usepackage{xcolor}
\usepackage{soul}
\usepackage{caption}

\title {Estimation and inference for causal spillover effects in egocentric-network randomized trials in the presence of network membership misclassification}

\date{}

\author[1,2]{\normalsize Ariel Chao}
\author[1,2]{\normalsize Donna Spiegelman}
\author[3]{\normalsize Ashley Buchanan}
\author[1]{\normalsize Laura Forastiere}

\affil[1]{Department of Biostatistics, Yale School of Public Health}
\affil[2]{Center for Methods in Implementation and Prevention Science, Yale School of Public Health}
\affil[3]{Department of Pharmacy Practice, College of Pharmacy, the University of Rhode Island}

\begin{document}

\maketitle

\section*{Abstract}
To leverage peer influence and increase population behavioral changes, behavioral interventions often rely on peer-based strategies. A common study design that assesses such strategies is the egocentric-network randomized trial (ENRT), in which those receiving the intervention are encouraged to disseminate information to their peers. The Average Spillover Effect (ASpE) measures the impact of the intervention on participants who do not receive it, but whose outcomes may be affected by others who do. The assessment of the ASpE relies on assumptions about, and correct measurement of, interference sets within which individuals may influence one another’s outcomes. It can be challenging to properly specify interference sets, such as networks in ENRTs, and when mismeasured, intervention effects estimated by existing methods will be biased. In HIV prevention studies where social networks play an important role in disease transmission, correcting ASpE estimates for bias due to network misclassification is critical for accurately evaluating the full impact of interventions. We combined measurement error and causal inference methods to bias-correct the ASpE estimate for network misclassification in ENRTs, when surrogate networks are recorded in place of true ones, and validation data that relate the misclassified to the true networks are available. We investigated finite sample properties of our methods in an extensive simulation study, and illustrated our methods in the HIV Prevention Trials Network (HPTN) 037 study.    

\newpage

\section{Introduction}
\label{s:intro}

Causal inference is often conducted under the potential outcomes framework, which usually rules out interference among individuals, that is, it assumes that individual potential outcomes depend only on the individual's exposure and not that of others, as articulated in the Stable Unit Treatment Value Assumption (SUTVA) \citep{rubin1980randomization, imbens2015causal, hernan2010causal}. However, in many settings, interference, or spillover, may be present. For example, in the HIV Prevention Trials Network (HPTN) 037 study \citep{latkin2009efficacy}, the intervention is assigned to drug-injection networks, and participants' engagement in HIV prevention behaviors can be influenced by their peers, which can then affect their transmission risk. Understanding social interactions that drive disease transmission is key for strengthening interventions. There is evidence that interference occurs in non-infectious disease interventions as well \citep{fletcher2017causal,desrosiers2020diffusion}. Public health interventions should exploit this feature to amplify their effectiveness. In particular, peer education interventions, for which individuals are trained on prevention of risky behaviors and encouraged to disseminate knowledge and behavioral change in their peer networks, have been shown to effectively increase HIV knowledge and reduce risk behaviors among both those who receive training and network members \citep{cai2008long, medley2009effectiveness}. 

A common study design for assessing the effect of peer-based interventions, and frequently used in HIV risk behavior research, is the egocentric-network randomized trial (ENRT), in which each index participant is randomly assigned to receive the intervention, and data on HIV knowledge and behavioral change is collected on both the index participants and members of their egocentric network \citep{friedman2008group, khan2011incarceration, friedman2008relative}. ENRTs are able to assess both the direct effect of receiving the intervention on indexes, as well as the spillover effect on network members. We refer to the latter as the Average Spillover Effect (ASpE). Causal inference in the presence of interference relies on the specification of an interference set for each individual, within which individuals may influence one another’s outcomes but not outside of them. For methods that assume network interference, whereby interference occurs through social networks, interference sets are defined by such connections and typically restricted to direct ones, known as the neighborhood interference assumption \citep{aronow2017estimating, tchetgen2017auto, ogburn2017vaccines, forastiere2020identification, forastiere2022estimating, lee2021estimating}. Additionally, it is typically assumed that an individual's potential outcomes depend on not only their own exposure, but also a function of the exposures of individuals in their interference set, for example, the proportion of those receiving the intervention. This function is commonly referred to as an ``exposure mapping function" that represents the mechanism through which interference occurs \citep{manski2013identification, sofrygin2016semi, aronow2017estimating, forastiere2020identification}. It is typically assumed that networks are correctly measured; however, investigators may not have accurate information for identifying true network ties, and networks collected at study baseline may change throughout the study period. In fact, it has been shown that improved methods are needed for ascertaining social connections in network-based studies \citep{young2018network, rudolph2017rural}. When interference sets are misspecified, we will show that the ASpE estimated by existing approaches is biased. Bias correction methods are crucial to produce valid impact assessment.

The problem of mismeasured exposure has been extensively studied in the statistical literature. For instance, the matrix and inverse matrix methods have long been available to correct for misclassified proportions and functions of proportions, such as risk differences and ratios \citep{barron1977effects, marshall1990validation, greenland1988variance, morrissey1999matrix}. Causal inference literature has recently begun to address misclassified spillover exposures, which can arise when either the exposure mapping function or the interference set is misspecified. Under misspecified exposure mapping, \citet{aronow2017estimating} showed that the Horvitz-Thompson estimator is unbiased for the overall effect between two exposure levels, and \citet{savje2021causal} showed that the estimator is consistent even if specification errors, defined as differences between the potential outcomes based on the full exposure vector in a network and those based on the exposure mapping function, are large, as long as the misspecification process is random. Interference sets are misspecified when the extent of interference is wrongly assumed, such as when neighborhood interference is assumed but in fact interference extends beyond first-order neighbors. \citet{leung2022causal} proposes methods for ``approximate neighborhood interference", under which exposures of network members further from the index have smaller, but potentially nonzero, effects on the index. While \citet{leung2022causal} addresses misspecified interference sets from wrongly assumed extent of interference based on K-degree neighbors, interference sets can also be misspecified from network mismeasurement, resulting in mismeasured spillover exposures even if the extent of interference were correctly specified.

In this paper, we address the misspecification of interference sets due to network mismeasurement, while assuming that interference occurs only in first-degree neighbors. A few papers have studied this issue. \citet{aronow2021spillover} showed through simulations that the bias of the Horvitz-Thompson estimator increases with the proportion of mismeasured network ties. \citet{bhattacharya2020causal} explored learning network structures using a score-based model selection algorithm, where first a fully connected chain graph is assumed, then edges are deleted to yield the best improvement in the pseudo-Bayesian Information Criterion score. \citet{egami2021spillover} showed that the average network-specific spillover effect is biased when interference may occur through multiple types of networks that may be only partially observed, and proposed sensitivity analysis methods to address concerns of interactions in unobserved networks violating partial interference. \citet{hardy2019estimating} used mixture models to model the distribution of the latent true exposure conditions, assuming prior models for the number of true connections, and estimated causal effects using the Expectation-Maximization algorithm. \citet{zhang2020spillovers} identified true networks when two network proxies are observed for each individual, one of which is assumed to be an instrumental variable for the true network, and the other is assumed to contain only one type of measurement error - either including no false connections while allowing missing ones, or vice versa. While these publications aim to learn the true network in order the estimate causal effects, which can be computationally intensive, we pursue a different approach where we augment the study design to include a validation study in which true networks are ascertained alongside the observed, error-prone ones, allowing us to empirically estimate the network misclassification process in order to bias-correct the point and interval estimates of the ASpE \citep{carroll2006measurement, buonaccorsi2010measurement, spiegelman2000estimation}. No previous paper has taken this approach.

Our methods are applicable to ENRTs, where network membership among network members may be misclassified. In Section 2, we present methods to bias-correct the point and interval estimates of the ASpE developed under a main study / validation study design, where the main study contains data on the outcome and observed networks of the study population, and the validation study is assumed to contain measurements of the true and observed networks. Section 3 presents a simulation study used to investigate finite sample properties of our methods. Section 4 presents an application of our methods to the HPTN 037 study, where a contamination study allowed us to determine the participants' true exposures.

\section{Methods}
\label{s:method}

\subsection{Notation, assumptions, and causal estimand}
\label{ss:notation}

In an ENRT, index participants are recruited and randomly assigned to the intervention, each nominating a set of network members. Let $ik$ be the $i$-th participant in the $k$-th network, with $i=1,\dots, n_k$ and $k=1,\dots,K$, for $N$ study participants. For ease of notation, let $i=1$ correspond to the index and $i=2,...,n_k$ correspond to network members for all $k$. We define an egocentric network $k$ to be the index $1k$ and their network members $ik$, with $i>1$.

Let $A_{ik}$ be a binary individual exposure status, where $A_{ik}=1$ if $ik$ receives the intervention and $0$ otherwise. In ENRTs, $A_{ik}=1$ for index $1k$ if $1k$ receives the intervention and 0 otherwise, whereas $A_{ik}=0$ for all network members $ik$, with $i>1$, $\forall k$. Let $\mathbf{H}_{N\times K}$ be an index matrix, where $H_{ik}=1$ if $ik$ is the $k$th index. Let $\mathbf{R}_{K\times 1}$ be the intervention allocation vector, where $R_k=1$ if network $k$ is randomized to intervention, such that $P_R=Pr(R_k=1)$ is the intervention allocation probability. Then, the individual exposure vector is given by $\mathbf{A}=\mathbf{H}\mathbf{R}$, assumed to be perfectly measured since we know the randomization design. 

To define potential outcomes, we first make the neighborhood interference assumption. Let $\mathcal{N}_{ik}$ be $ik$'s network neighborhood, comprising of individuals who share a network link with $ik$. Let $Y_{ik}(\mathbf{A})$ be $ik$'s potential outcome under the exposure vector $\mathbf{A}$, and $\mathbf{A}_{\mathcal{N}_{ik}}$ be the exposure vector in $ik$'s network neighborhood, $\mathcal{N}_{ik}$. We formally define the assumption as: 

\textbf{Assumption 1} (Neighborhood Interference). Let $g()$ be an exposure mapping function. For all $\mathbf{A}$, $\mathbf{A}'$ such that $A_{ik}=A_{ik'}$ and $g(\mathbf{A}_{\mathcal{N}_{ik}})=g(\mathbf{A'}_{\mathcal{N}_{ik}})$, then $Y_{ik}(\mathbf{A}) = Y_{ik}(\mathbf{A}')$. 

Under Assumption 1, $ik$'s potential outcome can be indexed by $A_{ik}$ and $G_{ik}$, where $G_{ik}=g(\mathbf{A}_{\mathcal{N}_{ik}})$, which we refer to as the spillover exposure, and $Y_{ik}(a,g)$ is thus the outcome that would be observed under $A_{ik}=a$ and $G_{ik}=g$. To define $G_{ik}$, we additionally assume that interference occurs through the number of treated first-order neighbors. Formally: 

\textbf{Assumption 2} (Exposure Mapping). Let $G_{ik}=\sum_{j\in \mathcal{N}_{ik}} A_{jk}$, then Assumption 1 holds.

We further assume non-overlapping networks, where indexes are not connected among themselves, and network members can only be connected to one index. Let $\mathbf{M}_{N\times K}$ be the true membership matrix, where $M_{ik}=1$ if $ik$ is a network member of index $k$, and $M_{ik}=0$ otherwise. Given $\mathbf{M}$, the network neighborhood for an index $1k$, $\mathcal{N}_{1k}$, is given by $\{ik: M_{ik}=1\}$. 
Then, formally:

\textbf{Assumption 3} (Non-overlapping Networks). 
\[M_{1k}=0 \quad \text{and} \quad \sum_k M_{ik} \le 1 \quad \forall k. \]

Because network members are typically not asked to delineate all of their network members in an ENRT, we may not observe their full network neighborhoods, which can include members in the same egocentric network or other egocentric networks, as well as out-of-study individuals. By assuming that $A_{ik}=0$ for every network member in the study and out-of-study individual, and that networks are non-overlapping (Assumption 3), $G_{ik}=1$ or $0$ for network members depending on the intervention received by their index, and is $0$ for index participants. As such, $\mathbf{G}=\mathbf{M}\mathbf{R}$, implying that $G_{ik}=0$ for all indexes, $G_{ik}=1$ for a network member in an intervention network, and $G_{ik}=0$ for a network member in a control network. Then, the set of possible potential outcomes for $ik$ are $Y_{ik}(a,g)\in \{Y_{ik}(1,0),Y_{ik}(0,1),Y_{ik}(0,0)\}$. Let $Y_{ik}$ be $ik$'s observed outcome. Outcomes are assumed to be binary throughout this paper. Although Assumptions 1-3 are not empirically verifiable, we assume that they are correct such that $G_{ik}$ is well-defined.

Our causal estimand, the ASpE, is the comparison of potential outcomes of network members if, possibly contrary to fact, all were exposed to the intervention through an index, versus not, while not receiving the intervention themselves. 
We consider the ASpE ($\delta$) as a risk difference (RD) as common in causal inference literature, and a risk ratio (RR) where effect modification is rare \citep{spiegelman2017evaluating}, defined as the following: 

\begin{equation}
    \delta_{RD}=\mathbb{E}[Y_{ik}(0,1)-Y_{ik}(0,0)|i>1]
\end{equation}

\begin{equation}
    \delta_{RR} = \frac{\mathbb{E}[Y_{ik}(0,1)|i>1]}{\mathbb{E}[Y_{ik}(0,0)|i>1]}
\end{equation}
where the expectation is taken over the distribution of potential outcomes of network members, assumed to be random variables \citep{hernan2010causal}.

Lastly, we make identification assumptions necessary for valid causal inference of the ASpE:

\textbf{Assumption 4} (Positivity) For $i>1$,
\[0 < Pr(A_{ik}=a,G_{ik}=g) < 1, \quad \forall (a,g) \in \{(0,1),(0,0)\}\]

\textbf{Assumption 5} (Unconfoundedness of the intervention) For $i>1$,
\[Y_{ik}(a, g) \perp A_{ik}, G_{ik}, \quad \forall (a,g) \in \{(0,1),(0,0)\}\]
where Assumptions 4 and 5 hold by the ENRT design. Given both assumptions, the ASpE can be identified as follows:

\begin{equation}
    \delta_{RD}=\mathbb{E}[Y_{ik}|A_{ik}=0,G_{ik}=1]-\mathbb{E}[Y_{ik}|A_{ik}=0,G_{ik}=0,i>1]
\end{equation}

\begin{equation}
    \delta_{RR}=\frac{\mathbb{E}[Y_{ik}|A_{ik}=0,G_{ik}=1]}{\mathbb{E}[Y_{ik}|A_{ik}=0,G_{ik}=0,i>1]}
\end{equation}
Then, an unbiased estimator of the right side quantities would be unbiased for $\delta_{RD}$ and $\delta_{RR}$.

Unlike $A_{ik}$, $G_{ik}$ can be misclassified due to network misclassification. Let $\mathbf{M^*}$ be the observed membership matrix, and $P_M = Pr(M_{ik}^*=1 | M_{ik}=1)$ be the probability of being classified into the correct network. Throughout this paper, $P_M$ is assumed to be independent of covariates; however, this assumption may be relaxed, and $P_M$ may be represented by a model that includes covariates. Let $G^*_{ik}$ denote the observed spillover exposure, and $\mathbf{G^*}=\mathbf{M^*}\mathbf{R}$ the observed spillover exposure vector. In the absence of $G_{ik}$ where only $G^*_{ik}$ is observed, we would instead estimate the following quantities, replacing $G_{ik}$ with $G^*_{ik}$:

\begin{equation}
    \delta^*_{RD}=\mathbb{E}[Y_{ik}|A_{ik}=0,G^*_{ik}=1]-\mathbb{E}[Y_{ik}|A_{ik}=0,G^*_{ik}=0,i>1]
\end{equation}

\begin{equation}
    \delta^*_{RR}=\frac{\mathbb{E}[Y_{ik}|A_{ik}=0,G^*_{ik}=1]}{\mathbb{E}[Y_{ik}|A_{ik}=0,G^*_{ik}=0,i>1]}
\end{equation}
The relationship between $\delta$ and $\delta^*$ will be derived in Section~\ref{ss:bias.analysis} under the assumption of non-differential misclassification, which assumes that the measurement error process is independent of potential outcomes conditional on the true spillover exposure. Formally, 

\textbf{Assumption 6} (Non-differential misclassification)
\[Y_{ik}(a,g) \perp G_{ik}^*|G_{ik}, \quad i>1.\]

\subsection{Main study: estimation of the ASpE under misclassified spillover exposure}
\label{ss:main}

Under Assumption 5, the ASpE can be estimated using sample average estimators. If the true spillover exposure were observed, the ASpE would be estimated as follows: 
\begin{table}[h!]
\centering
\caption{$2\times 2$ table for the estimation of the ASpE under the true, unobserved exposure for $i>1$} \label{t:tableone}
{\begin{tabular*}{\textwidth}{@{}l@{\extracolsep{\fill}}c@{\extracolsep{\fill}}c
@{\extracolsep{\fill}}c@{}}
\hline & $G_{ik}$=1 & $G_{ik}$=0 &  \\
    \hline
    $Y_{ik}$=1 & $\mathpzc{a}$ & $\mathpzc{b}$ & $m_1$\\
    $Y_{ik}$=0 & $\mathpzc{c}$ & $\mathpzc{d}$ & $m_0$\\
    & $n_1$ & $n_0$ & $N$ \\ \hline 
 \end{tabular*}
\vspace*{-6pt}}
\end{table}
\[ \widehat{\delta}_{RD} = \frac{\mathpzc{a}}{n_1} - \frac{\mathpzc{b}}{n_0} \] 
\[ \widehat{\delta}_{RR} = \frac{\mathpzc{a}}{n_1} / \frac{\mathpzc{b}}{n_0} \]
where data summaries $\mathpzc{a}$, $\mathpzc{b}$, $n_1$, and $n_0$, can be found in Table~\ref{t:tableone}. These estimators are unbiased for the quantities on the right hand side of the identifying formulas (3) and (4), and therefore, under Assumption 5, are unbiased for $\delta_{RD}$ and $\delta_{RR}$.

When instead the true spillover exposure is not observed, estimates must rely on statistics based on the observed networks, such as the ones reported in Table~\ref{t:tabletwo}. Here, upper script letters are used to denote possibly misclassified data summaries under the observed spillover exposure. In this case, the ASpE estimators would be as follows:
\begin{table}[h!]
\centering
\caption{$2\times 2$ table for the estimation of the ASpE under the observed and possibly misclassified exposure for $i>1$} \label{t:tabletwo}
{\begin{tabular*}{\textwidth}{@{}l@{\extracolsep{\fill}}c@{\extracolsep{\fill}}c
@{\extracolsep{\fill}}c@{}}
\hline & $G^*_{ik}$=1 & $G^*_{ik}$=0 &  \\
    \hline
    $Y_{ik}$=1 & $\mathpzc{A}$ & $\mathpzc{B}$ & $m_1$\\
    $Y_{ik}$=0 & $\mathpzc{C}$ & $\mathpzc{D}$ & $m_0$\\
     & $\mathpzc{N}_1$ & $\mathpzc{N}_0$ & $N$ \\ \hline 
 \end{tabular*}
\vspace*{-6pt}}
\end{table}
\[ \widehat{\delta}_{RD}^* = \frac{\mathpzc{A}}{\mathpzc{N}_1} - \frac{\mathpzc{B}}{\mathpzc{N}_0} \]
\[ \widehat{\delta}_{RR}^* = \frac{\mathpzc{A}}{\mathpzc{N}_1} / \frac{\mathpzc{B}}{\mathpzc{N}_0} \] 
These estimators are unbiased for the quantities on the right hand side of the identifying formulas in (5) and (6), and therefore, under Assumption 5, are unbiased for $\delta_{RD}^*$ and $\delta_{RR}^*$.

\subsection{Bias analysis of $\widehat{\delta}_{RD}^*$ and $\widehat{\delta}_{RR}^*$}
\label{ss:bias.analysis}

Define the outcome rate for $i>1$ as $P_{Y0} \times \delta_{RR}^{\quad G_{ik}}$, where $P_{Y0}$ is the baseline outcome rate when $A_{ik}=0$ and $G_{ik}=0$, such that the outcome rate is $P_{Y0}\delta_{RR}$ for $G_{ik}=1$ and $P_{Y0}$ for $G_{ik}=0$. The true risk difference is thus given by $\delta_{RD}=P_{Y0}\delta_{RR}-P_{Y0}=P_{Y0}(\delta_{RR}-1)$. Under misclassified spillover exposures, $\widehat{\delta}_{RD}$ and $\widehat{\delta}_{RR}$ are biased, and the relationship between $\delta^*$ and $\delta$ is as follows under Assumption 6:

\[\delta_{RD}^* = P_{Y0}P_M(\delta_{RR}-1), \quad \text{and}\]
\[\delta_{RR}^* = \frac{[\delta_{RR}\{P_M+(1-P_M)P_R\} + (1-P_M)(1-P_R)](1-P_R)}{P_R \delta_{RR}[1-\{P_M + (1-P_M)P_R\}] + \{P_M + (1-P_M)(1-P_R)\}(1-P_R)} \vspace{0.5cm}\]
Derivations are provided in Appendix 2. Given $\delta_{RD}=P_{Y0}(\delta_{RR}-1)$ and $\delta_{RD}^*=P_{Y0}P_M(\delta_{RR}-1)$, it is easily seen that $\delta^*_{RD}\le \delta_{RD}$ and therefore always biased towards the null under the non-differential misclassification assumption, except when $P_M=1$ such that $\delta^*_{RD}=\delta_{RD}$. The same can be shown for $\delta^*_{RR}$. The bias and absolute relative (Rel.) bias with respect to the true effect, are:
\[\text{Bias}_{\widehat{\delta}_{RD}^*}  = P_{Y0}(\delta_{RR}-1)(P_M-1)\]
\[\text{Rel. Bias}_{\widehat{\delta}_{RD}^*} = 1-P_M\]
\[\text{Bias}_{\widehat{\delta}_{RR}^*} = \frac{[\delta_{RR}\{P_M+(1-P_M)P_R\} + (1-P_M)(1-P_R)](1-P_R)}{P_R \delta_{RR}[1-\{P_M + (1-P_M)P_R\}] + \{P_M + (1-P_M)(1-P_R)\}(1-P_R)} - \delta_{RR}\]
\begin{align*}
\text{Rel. Bias}_{\widehat{\delta}_{RR}^*} &= \displaystyle\left\lvert \frac{\{\delta_{RR}[P_M+(1-P_M)P_R] + (1-P_M)(1-P_R)\}(1-P_R)}{\delta_{RR}\{P_R \delta_{RR}[1-(P_M + (1-P_M)P_R)] + [P_M + (1-P_M)(1-P_R)](1-P_R)\}} -1 \right\rvert
\end{align*} 

\vspace{0.5cm}
\noindent where there the bias is zero when $P_M=1$. The behavior of bias as a function of each parameter is described in Section~\ref{ss:simres}, and derivations are provided in Appendix 3.

\subsection{Validation study: estimation of network misclassification parameters}
\label{ss:validation}

When $G^*_{ik} \ne G_{ik}$ as a result of network misclassification, the estimated ASpE is biased. However, when data on true networks or spillover exposures are available in a subpopulation, we can extend methods developed under a main study / validation study approach to bias-correct the ASpE estimated under mismeasured spillover exposures \citep{barron1977effects, marshall1990validation, spiegelman2000estimation}. In the validation study, the relationship between the true and observed spillover exposure is modeled. Here, we estimate the sensitivity ($\theta$) and specificity ($\phi$) of spillover exposure classification among network members, to be used to bias-correct the ASpE estimated in the main study. We can define $\theta$ and $\phi$ as functions of $P_M$ and $P_R$. Intuitively, if $ik$ were exposed to a treated index, they could either truly belong to their assigned intervention network $k$, or a different network $k'$ that is also an intervention network. An analogous argument can be made if $ik$ were unexposed to a treated index. Therefore, $\theta$, and $\phi$ are defined as follows: 
\[\theta = Pr(G_{ik}^*=1|G_{ik}=1) = P_M + (1-P_M)P_R, \; \text{and} \vspace{-0.3cm}\]
\[\phi = Pr(G_{ik}^*=0|G_{ik}=0) = P_M + (1-P_M)(1-P_R), \quad i>1 \] 
where it is easily seen that $\theta=\phi$ under $P_R=0.5$ as is often the case. Proofs of the above relationships are provided in Appendix 1. When data on true networks are available in the validation study such that $P_M$ is estimable, and given that $P_R$ is known in an ENRT, $\theta$ and $\phi$ can be estimated by these formulas. If data are only available for true spillover exposures but not true networks, and $P_M$ cannot be estimated, assuming that network misclassification does not vary by covariates, a simple $2 \times 2$ table shown in Table~\ref{t:tablethree} can be used to estimate $\theta$ and $\phi$ among $i>1$ using their true and observed spillover exposures:
\begin{table}[h!]
\centering
\caption{$2\times 2$ table for the estimation of $\theta$ and $\phi$ in the validation study} \label{t:tablethree}
{\begin{tabular*}{\textwidth}{@{}l@{\extracolsep{\fill}}c@{\extracolsep{\fill}}c
@{\extracolsep{\fill}}c@{}}
\hline & $G_{ik}$=1 & $G_{ik}$=0 &  \\
    \hline
    $G^*_{ik}$=1 & $n_{v11}$ & $n_{v01}$ & $n_{v.1}$\\
    $G^*_{ik}$=0 & $n_{v10}$ & $n_{v00}$ & $n_{v.0}$\\
     & $n_{v1.}$ & $n_{v0.}$ & $n_{v}$ \\ \hline 
 \end{tabular*}
\vspace*{-6pt}}
\end{table}
\[\vspace{-0.3cm} \widehat{\theta} = \frac{n_{v11}}{n_{v1.}}, \; \text{and} \quad \widehat{\phi} = \frac{n_{v00}}{n_{v0.}}  \vspace{0.5cm}\]

Besides $\theta$ and $\phi$, other misclassification parameters can also be estimated in the validation study, such as the positive and negative predictive values, $PPV=Pr(G_{ik}=1|G^*_{ik}=1)$ and $NPV=Pr(G_{ik}=0|G^*_{ik}=0)$, respectively. The usage of $\widehat{\theta}$, $\widehat{\phi}$, $\widehat{PPV}$, and $\widehat{NPV}$ for bias-correction will be further discussed in Section~\ref{ss:bias.correction}. 

Note that a validation study can be internal or external to the main study; it is internal when conducted on a subset of the main study, as is the case in our illustrative example, HPTN 037, and external when conducted on individuals outside of the main study. We assume that measurement error process estimated in the validation study is generalizable to the main study, such that the parameters of the misclassification model needed to correct effect estimates obtained from the main study are the same between the main study and validation study. Formally,

\textbf{Assumption 7} (Generalizability) Let $V_{ik}$ be an indicator for whether $ik$ is included in the internal validation study. Then, $G^*_{ik} \perp V_{ik} | G_{ik}$.

When an external validation study is used, a similar assumption of transportability must be invoked. This assumes that the parameters necessary for correction of bias due to misclassification that generated the validation study are the same as those that generated the main study data, although the latter are not estimable \citep{carroll2006measurement}.

\subsection{Methods for bias correction}
\label{ss:bias.correction}

\subsubsection{Matrix method estimator}
\label{sss:matrix}

As shown in Section \ref{ss:main}, because the ASpE under the observed spillover exposure is estimated with misclassified counts, one approach to bias correction is to bias-correct these counts by estimating their true counterparts. Given the binary nature of $G_{ik}$, we extend the matrix method \citep{barron1977effects}, developed under Assumption 6. This assumption is empirically verifiable in a main study / internal validation study design under Assumption 5 when the correlation of $Y_{ik},G^*_{ik}|G_{ik}$ for $i>1$ is near 0.
The matrix method additionally assumes correctly measured outcomes, such that $\mathpzc{A}+\mathpzc{B}=a+b$. Given $\widehat{\theta}$ and $\widehat{\phi}$ from the validation study, the bias-corrected estimators are given by: \vspace{0.5cm}
\[\widehat{\delta}_{RD} = \frac{\mathpzc{B}-\widehat{\phi}m_1}{N_0-\widehat{\phi}N} -\frac{\mathpzc{A}-\widehat{\theta}m_1}{N_1-\widehat{\theta}N} \]
\[\widehat{\delta}_{RR} =\frac{(\mathpzc{B}-\widehat{\phi}m_1)(N_1-\widehat{\theta}N)}{(\mathpzc{A}-\widehat{\theta}m_1)(N_0-\widehat{\phi}N)} \]

\vspace{0.5cm}
\noindent Derivations are provided in Appendix 4. 

Under misclassified exposures, the multivariate delta method is used to derive the asymptotic variance of $\widehat{\delta}_{RD}^*$ and $\widehat{\delta}_{RR}^*$, using the standard binomial variance for the binary outcomes, and ignoring the covariance between $Y_{ik}(0,1)$ and $Y_{ik}(0,0)$ \citep{neyman1923application, imbens2015causal}. After applying our bias-correction method to obtain $\widehat{\delta}_{RD}$ and $\widehat{\delta}_{RR}$, variance estimation is provided by the multivariate delta method previously given by \citet{greenland1988variance}. However, these estimators do not consider the clustering of outcomes within networks, as would be expected in ENRTs. For example, in HPTN 037, the intracluster correlation coefficient (ICC) was 0.16. Therefore, to take this correlation into account, we propose two approaches to adjust the variance estimates for clustering. First, the variance estimate can be inflated by the design effect, $1+(\bar{m}-1)\rho$, where $\bar{m}$ is the average network size, and $\rho$ is the estimated ICC \citep{donner2000design}. Second, the variance can be estimated by network bootstrapping where both the validation and main study networks are randomly resampled as a whole to maintain the within-network correlation structure, enabling us to obtain estimates that appropriately account for clustering \citep{davison1997bootstrap}. 

To note, for good finite-sample performance of the bias-correction estimators, the following constraints should be considered:
\begin{enumerate}
    \item $\theta+\phi \ne 1$. When $\theta+\phi=1$, the misclassification matrix (Appendix 4) is non-invertible since the determinant, $\frac{1}{\theta+\phi-1}=0$, giving undefined cell counts.  
    \item When $\theta+\phi>1$, to avoid estimating negative cell counts $\widehat{\mathpzc{a}}$, $\widehat{\mathpzc{b}}$, $\widehat{n_1}$, and $\widehat{n_0}$:
    \[\theta = Pr(G_{ik}^*1|G_{ik}=1) > \text{max}\left(Pr(G^*_{ik}=1|Y_{ik}=1),Pr(G^*_{ik}=1|Y_{ik}=0)\right), \quad \text{and} \vspace{-0.3cm}\]
    \[\phi=Pr(G_{ik}^*=0|G_{ik}=0) > \text{max}\left(Pr(G^*_{ik}=0|Y_{ik}=1),Pr(G^*_{ik}=0|Y_{ik}=0)\right).\] 
    \item When $\theta+\phi<1$, to avoid estimating negative cell counts $\widehat{\mathpzc{a}}$, $\widehat{\mathpzc{b}}$, $\widehat{n_1}$, and $\widehat{n_0}$, 
    \[\theta=Pr(G_{ik}^*=1| G_{ik}=1) < \text{min}\left(Pr(G^*_{ik}=1|Y_{ik}=1),Pr(G^*_{ik}=1|Y_{ik}=0)\right), \quad \text{and} \vspace{-0.3cm}\] 
    \[\phi=Pr(G_{ik}^*=0|G_{ik}=0) < \text{min}\left(Pr(G^*_{ik}=0|Y_{ik}=1),Pr(G^*_{ik}=0|Y_{ik}=0)\right).\]
\end{enumerate}
Details about the restrictions are provided in Appendix 5. 

\subsubsection{Inverse matrix method estimator}
\label{sss:inv.matrix}
If the above constraints are not met, the corrected ASpE estimate will be inaccurate. In this case, the inverse matrix method \citep{marshall1990validation,greenland1988variance}, which does not impose these restrictions, may be considered; however, the matrix method estimator gains efficiency relative to the inverse matrix estimator when the outcome is rare and the validation study is small \citep{morrissey1999matrix}, as is the case in our illustrative example, HPTN 037. The inverse matrix method corrects effect estimates using $PPV$ and $NPV$ estimated separately for cases ($Y_{ik}=1$) and non-cases ($Y_{ik}=0$), with the following estimators: 
\begin{align*}
    \widehat{\delta}_{RD}^{IM}&=\frac{\widehat{PPV}_1\mathpzc{A}+(1-\widehat{NPV}_1)\mathpzc{B}}{\widehat{PPV}_1\mathpzc{A}+(1-\widehat{NPV}_1)\mathpzc{B}+\widehat{PPV}_0\mathpzc{C}+(1-\widehat{NPV}_0)\mathpzc{D}} \\
    &\quad -\frac{(1-\widehat{PPV}_1)\mathpzc{A}+\widehat{NPV}_1\mathpzc{B}}{(1-\widehat{PPV}_1)\mathpzc{A}+\widehat{NPV}_1\mathpzc{B}+(1-\widehat{PPV}_0)\mathpzc{C}+\widehat{NPV}_0\mathpzc{D}}
\end{align*}

\[\widehat{\delta}_{RR}^{IM}=\frac{\{\widehat{PPV}_1\mathpzc{A}+(1-\widehat{NPV}_1)\mathpzc{B}\}\{(1-\widehat{PPV}_1)\mathpzc{A}+\widehat{NPV}_1\mathpzc{B}+(1-\widehat{PPV}_0)\mathpzc{C}+\widehat{NPV}_0\mathpzc{D}\}}{\{(1-\widehat{PPV}_1)\mathpzc{A}+\widehat{NPV}_1\mathpzc{B}\}\{\widehat{PPV}_1\mathpzc{A}+(1-\widehat{NPV}_1)\mathpzc{B}+\widehat{PPV}_0\mathpzc{C}+(1-\widehat{NPV}_0)\mathpzc{D}\}}\]

\vspace{0.5cm}
\noindent where the subscript for $PPV$ and $NPV$ indicates the estimation among cases or non-cases. Full derivations are provided in Appendix 6. Variance estimation of the inverse matrix method estimators can also be achieved as described in Section~\ref{sss:matrix}.

\subsubsection{Maximum likelihood estimator (MLE)}
While the matrix and inverse matrix estimators are easily implementable, there is no clear way to directly incorporate the effect of clustering. As such, we also propose a maximum likelihood method to allow for this clustering. We can construct an observed data likelihood using a mixed effects model. Let $Pr(Y_{ik}=1)=P_{Y0}+\delta_{RD} G_{ik}+b_k$ be a model that estimates $\delta_{RD}$, where $b_k \sim N(0,\sigma^2_B)$ is the network random effect. Let $V_{ik}$ be an indicator for being in the internal validation study, and $\boldsymbol\omega=(P_{Y0},\delta_{RD},\sigma^2_B, PPV, NPV)^T$. Following \citet{spiegelman2000estimation} and \cite{zhou2020maximum}, assuming no clustering of the misclassification process, the likelihood function is given by:

\begin{align*}
    L(\boldsymbol\omega) &= \prod_{k=1}^K\int \left\{\prod_{i \in \{i>1, V_{ik}=0\}}^{n_k}\sum_{g=0}^1\left(P_{Y0}+\delta_{RD} g+b_k\right)^{Y_{ik}}\left(1-P_{Y0}-\delta_{RD} g-b_k\right)^{1-Y_{ik}}Pr(G_{ik}=g|G^*_{ik})\right.\\
    &\left.\quad \times \prod_{i \in \{i>1, V_{ik}=1\}}^{n_k} \left(P_{Y0}+\delta_{RD} G_{ik}+b_k\right)^{Y_{ik}}\left(1-P_{Y0}-\delta_{RD} G_{ik}-b_k\right)^{1-Y_{ik}}\right\} f(b_k) db_k\\
    &\quad \times \prod_{i \in \{i>1, V_{ik}=1\}}^{n_k} Pr(G_{ik}=0|G_{ik}^*=0)^{(1-G_{ik})(1-G_{ik}^*)}[1-Pr(G_{ik}=0|G_{ik}^*=0)]^{G_{ik}(1-G_{ik}^*)}\\
    &\quad \times [1-Pr(G_{ik}=1|G_{ik}^*=1)]^{(1-G_{ik})G_{ik}^*}Pr(G_{ik}=1|G_{ik}^*=1)^{G_{ik}G_{ik}^*}
\end{align*}
Note that the second product term in the integral is excluded for a main study / external validation study design, and the limits of integration over $b_k$ are imposed to ensure that probabilities are $\in (0,1)$. For the estimation of $\delta_{RR}$, we replace the linear model with a log-binomial model in the construction of the likelihood. The variance can be estimated by the inverse of the observed information matrix under the observed likelihood.

\section{Simulation study}
\label{s:sim}

\subsection{Design}
\label{ss:simdes}

Based on a real-world ENRT, HPTN 037 \citep{latkin2009efficacy}, we conducted a simulation study under scenarios of varying values of $P_{Y0}$, $\delta_{RR}$, $P_M$, and $P_R$ to assess trends in bias and confirm that empirical results align with those derived analytically for both the matrix and inverse matrix methods. The simulation sample consisted of $K=1000$ with $n_k=3$ for all $k$, for a total of $2000$ network members. Under $P_M \in$ (0.1 0.25, 0.5, 0.75, 0.9), $P_{Y0} \in$ (0.1, 0.25, 0.5, 0.75, 0.9), $\delta_{RR} \in$ (0.25, 0.75, 1.25, 3, 5), and $P_R \in$ (0.2, 0.5, 0.8), a total of 255 scenarios were studied, excluding 120 scenarios with combinations of $P_{Y0}$ and $\delta_{RR}$ that gave $P_{Y0}\delta_{RR}>1$. Outcomes were additionally simulated to be correlated within networks to assess the robustness of the estimators to such correlations under a model for $i>1$, $Pr(Y_{ik}=1) \sim expit\{P_{y0}+\delta_{RR}G_{ik}+b_k\}$ where $b_k \sim N(0,\sigma_b^2)$,
$ICC=\frac{\sigma^2_b}{\sigma^2_b+\sigma^2_e}$ such that $\sigma^2_b=\frac{\sigma^2_e ICC}{(1-ICC)}$ for $ICC \in$ (0.005, 0.01, 0.1, 0.25), and $\sigma_e^2=3.29$ is set to the standard logistic regression residual variance. Each network member was correctly classified with probability $P_M$, and if misclassified, each network except the true network had equal probability of being the misclassified network. Each scenario was simulated 2000 times, where the bias-corrected ASpE estimates were taken as empirical averages across all replications and compared to the true values, and empirical standard errors were compared to the theoretical ones. 

\subsection{Results}
\label{ss:simres}

Under each scenario, the empirical bias of the ASpE was virtually identical to the theoretical one, so was the ASpE estimated with the matrix method estimator to the true value in most scenarios, except in those where the sum of $\theta$ and $\phi$ was close to 1 and thereby causing the misclassification matrix to be unstable, a limitation mentioned in Section~\ref{ss:bias.correction}. However, in these scenarios, the inverse matrix method estimators performed well. When outcomes were correlated, the point estimates were accurately bias-corrected, while the empirical standard errors were slightly larger than the theoretical estimates in some scenarios, with more noticeable differences observed with higher ICCs. The empirical standard errors in 35\% of the scenarios were larger than their theoretical counterparts by greater than $10\%$ under $ICC=0.25$, while this was the case for only 17\% of scenarios under $ICC=0.10$, and a simple adjustment of the theoretical estimates using the design effect inflation was able to achieve comparability to the empirical ones. Although our variance estimator appears fairly robust to such correlations in ENRTs, variance adjustment for clustering is recommended, especially when ICC is high. 

The mean and relative empirical bias of $\widehat{\delta}^*_{RD}$ and $\widehat{\delta}^*_{RR}$ by $1-P_M$, $P_{Y0}$, $\delta_{RR}$, and $P_R$, are presented in Appendix Tables 1-4, respectively, each averaged over the other parameters. The bias estimates under all scenarios considered had 95\% coverage probability on average. As expected from the bias formulae given in Section~\ref{ss:bias.analysis}, we found that the ASpE estimated under misclassified networks was biased towards the null, with the strength of the bias dependent on the value of these parameters. For $\widehat{\delta}^*_{RD}$, the magnitude of the bias did not depend on $P_R$ and was linear in all parameters, as shown in Section~\ref{ss:bias.analysis}. With constant $P_{Y0}$ and $\delta_{RR}$, the bias of $\widehat{\delta}^*_{RD}$ increased as the misclassification probability increased. With constant $P_M$ and $\delta_{RR}$, the bias of $\widehat{\delta}^*_{RD}$ increased as $P_{Y0}$ increased. With constant $P_{Y0}$ and $P_M$, the bias of $\widehat{\delta}^*_{RD}$ increased as $\delta_{RR}$ diverged from 1. For $\widehat{\delta}^*_{RR}$, the bias did not depend on $P_{Y0}$ and was no longer linear in the other parameters. With constant $P_R$ and $\delta_{RR}$, the bias of $\widehat{\delta}^*_{RR}$ increased as the misclassification probability increased. With constant $P_M$ and $\delta_{RR}$, the bias of $\widehat{\delta}^*_{RR}$ decreased slightly but stayed relatively constant as $P_R$ increased. Lastly, with constant $P_R$ and $P_M$, the bias of $\widehat{\delta}^*_{RR}$ increased as $\delta_{RR}$ diverged from 1.

\section{Illustrative example: HPTN 037}
\label{s:ill}

HPTN 037 was a phase III ENRT conducted in the Philadelphia, Pennsylvania, and Chiang Mai, Thailand, that assessed the impact of a network-oriented peer education program on promoting HIV risk reduction behaviors among people who inject drugs \citep{latkin2009efficacy}. The study recruited 414 index participants who were asked to recruit at least one drug or sex network member to form their network neighborhoods. Indexes were randomized to intervention or control networks, where those in intervention networks received peer-educator training to encourage injection and sexual risk reduction with their network members. The primary study found that the number of participants reporting injection risk behaviors dropped significantly between baseline and follow-up in both arms, with the intervention arm having a larger decrease. Here, we included only participants at the Philadelphia site, as the ``war on drugs" in Thailand during the study period might have altered interactions among people who injected drugs \citep{vongchak2005influence}. We focused on a composite outcome defined as the occurrence of any reported HIV risk behavior, such as sharing injection equipment, at the 12-month visit. Buchanan et al. also estimated spillover effects of the intervention on this outcome using generalized estimating equations and marginal structural models, assuming correctly classified networks \citep{buchanan2018assessing, buchanan2022assessing}.

Here, $A_{ik}$ is the individual intervention assignment and is 0 for all network members, and $G^*_{ik}$ is defined as the observed spillover exposure to the intervention, as the study was previously analyzed \citep{latkin2009efficacy}. To apply our methods, $G_{ik}$ can be measured either by direct observation of the true networks, or through other forms of validation data. Here, $G_{ik}$ is defined using the study's exposure contamination survey \citep{aroke2022evaluating}. At the 6-month follow-up visit, each participant was asked to recall terminology associated with the intervention training; if any of five intervention-associated terms were recalled, this was taken as evidence of intervention exposure. Specifically, if a network member from either arm recalled any of the five terms, we assumed that they were exposed to the intervention through a treated index such that $G_{ik}=1$; likewise, if none of the five terms were recalled, we assume that there was no exposure from an index and that $G_{ik}=0$. Participants were also asked to recall a positive control term, to which everyone was exposed regardless of network intervention status, as well as three negative control terms unrelated to the study intervention that should not have been recalled. Because some participants might have been exposed to the intervention but not recall the terms specifically, or might have said they knew the terms due to social desirability, only network members who recalled the positive control term and none of the negative control terms were included in the validation study for greater accuracy in the modelling of the misclassification process. 

At the 12-month follow-up visit, the main study included 269 network members from 184 networks followed between December 2002 to July 2006, with 38 network members from 35 networks also included in the validation study. $\widehat{\theta}$ and $\widehat{\phi}$ were 0.60 and 0.79, respectively. There was no evidence of differential misclassification, as $\widehat{\theta}$ and $\widehat{\phi}$ estimated among cases and non-cases were not significantly different (p=0.92). The conditions from Section~\ref{ss:bias.correction} for using the matrix method were satisfied; here, $\theta$ and $\phi$ needed to be, and were, greater than max$(Pr(G^*=1|Y=1),Pr(G^*=1|Y=0))=0.53$ and max$(Pr(G^*=0|Y=1),Pr(G^*=0|Y=0))=0.65$ to ensure positive cell count estimates. The inverse matrix method may not be appropriate here because of the small validation study sample size as well as number of cases within in, such that the $PPV$ and $NPV$ to be estimated separately for cases and non-cases may not be accurate and adversely affect the accuracy of the bias-corrected estimates.

Table~\ref{t:tablefour} presents the $\widehat{\delta}_{RD}$ and $\widehat{\delta}_{RR}$ estimated under the observed spillover exposures and using our bias-correction estimators. The intervention had a significant spillover effect on the occurrence of risk behaviours. The ASpE was considerably biased towards the null due to substantial network misclassification: there was $66\%$ bias in the risk difference, and $124\%$ bias in the risk ratio. Several confidence intervals (CIs) are reported. First, without considering clustering, we estimated the variance using the multivariate delta method, which when accounting for the estimation of $\theta$ and $\phi$ using a small validation study sample size, was substantially inflated; however, the CI considerably narrowed when the validation study sample was artificially increased ten-fold. To take clustering into account, we inflated the variance by the design effect with an estimated ICC of 0.16 under a mixed effects model, performed network bootstrapping with 1000 bootstrap samples, and obtained the likelihood-based estimates. Here, we found that the variance estimated using the delta method without adjusting for clustering was robust to correlated outcomes, as it was comparable to the likelihood-based and bootstrap variances.

\begin{table}[h!]
\centering
\caption{Comparison of ASpE estimates and 95\% CIs from HPTN 037 under $\widehat{\theta}=0.60$ and $\widehat{\phi}=0.79$} \label{t:tablefour}
{\begin{tabular*}{\textwidth}{p{4.5cm}p{6cm}p{6cm}}
\hline & $\widehat{\delta}_{RD}$ & $\widehat{\delta}_{RR}$  \\
    \hline
    ITT estimate (95\% CI) & -0.15 (-0.26, -0.04) & 0.60 (0.41, 0.89)\\
    Bias-corrected estimate (95\% CI delta method, $n_v=38$) ($n_v=380$) & -0.44 (-1.25, 0.36) (-0.84, -0.05) & 0.27 (0.05, 1.38) (0.09, 0.78) \\
    Bias-corrected estimate (95\% CI, design effect inflation with $\widehat{ICC}$ = 0.16, $n_v$ = 38) ($n_v=380$) & -0.44 (-0.34, 0.45) (-0.89, -0.01) & 0.27 (0.04, 1.67) (0.08, 0.88) \\
    Bias-corrected estimate (95\% CI, bootstrap variance with resampling of both validation and main study networks) & -0.44 (-0.80, -0.08) & 0.27 (0.08, 0.87)\\
    MLE (95\% CI, $n_v=38$)  & -0.43 (-0.73, -0.13) & 0.26 (0.09, 0.73)\\
    \hline 
 \end{tabular*}
\vspace*{-6pt}}
\end{table}

\vspace{-1cm}
\section{Discussion}
\label{s:discuss}

Implementation studies assess how to best translate and scale up research evidence into public health practice. The impact of interventions, such as their reach and societal benefit, can be increased by leveraging behavioral spillover, especially in settings where resources are limited \citep{fletcher2017causal, desrosiers2020diffusion}. The assessment of the spillover effect can be challenging because in practice there is often substantial uncertainty about true network structures \citep{bhattacharya2020causal}. In this paper, we analyzed the bias of the ASpE estimate in an ENRT setting when the recorded networks are possibly misclassified, and developed bias-corrected estimators. Both the simulation study and application to real data showed that the ASpE estimate is biased towards the null under misclassified networks. This finding is in line with past statistical literature that showed misspecified exposures due to intervention noncompliance biased the estimated intervention effects toward the null hypothesis \citep{hernan2012beyond}.

In this paper, we assumed that networks were egocentric and non-overlapping. By making these assumptions, it is implied that (i) network members can only belong to one network; and (ii) index participants from different networks cannot interfere with one another. We can in principle relax (i), such that network members may be influenced by more than one index, by defining the spillover exposure to be the connection to at least one treated index, $G_{ik} = \mathbb{I}(\sum_{j \in \mathcal{N}_{ik}}{A_{jk}} > 0)$, and still apply our methods in the same way. We can further relax (ii), such that index participants can be exposed to the intervention from other indexes, and the potential outcome $Y_{ik}(1,1)$, which was previously undefined, is now identifiable given that we can now have $G_{ik}=1$ for some index members. Other causal effects, such as the total effect comparing potential outcomes if one were to receive both the individual intervention and spillover exposure versus having neither, $\mathbb{E}[Y_{ik}(1,1)]-\mathbb{E}[Y_{ik}(0,0)]$, can now be estimated. 

Our methods have a few limitations. Here, we focused on non-parametric methods applicable to an ENRT design, where both the exposure and outcome are binary, and no covariate adjustment is considered because intervention is randomized. While it is reasonable to assume that both observed and unobserved covariates are balanced on average under randomization, covariate adjustment may still be needed. A likelihood-based approach to estimate the ASpE and misclassification parameters, as detailed in Section~\ref{ss:bias.correction}, can include covariates, although this may be more computationally intensive. Furthermore, our methods rely on the neighborhood interference assumption; however, it is also possible that after the network members receive the education through their index, they may influence peers in other networks or even reinforce the knowledge of their index \citep{latkin2013dynamic}. Future methods will be developed for more complex forms of interference assumptions in the setting of CRTs, where multiple individuals per cluster are assigned to intervention, and other causal effects besides the spillover effect can be mismeasured. 

\newpage

\bibliographystyle{biom} \bibliography{main}

\newpage
\section*{Appendix}

\subsection*{Appendix 1: Derivation of $\theta$ and $\phi$ in the validation study}

For $ik$, whose observed network is $k$, $k$ could be the true network, or could be the misclassified network when the true network is $k'$. 

\begin{flalign*}
\theta &= Pr(G_{ik}^*=1|G_{ik}=1) \\
& = Pr(M_{ik}^*=1, R_k=1 | M_{ik}=1, R_k=1) + \underset{k\in \mathbf{K}_{-k'}}{\sum}Pr(M_{ik}^*=1, R_{k}=1 | M_{ik'}=1, R_{k'}=1) \\
& = Pr(M_{ik}^*=1| M_{ik}=1, R_k=1) \\
&\quad +\underset{k\in \mathbf{K}_{-k'}}{\sum} Pr(M_{ik}^*=1| M_{ik'}=1, R_{k'}=1)Pr(R_{k}=1|M_{ik'}=1, R_{k'}=1) \\
& = Pr(M_{ik}^*=1|M_{ik}=1) + \underset{k\in \mathbf{K}_{-k'}}{\sum}Pr(M_{ik}^*=1|M_{ik'}=1)Pr(R_{k}=1) \\
& = P_M + (1-P_M)P_R &%
\end{flalign*}

\noindent The derivation of $\theta$ and $\phi$ relies on the following assumptions:
\begin{itemize}
    \item $\mathbf{G^*}=\mathbf{M^*R}$. $Pr(M_{ik}^*=1, R_k=1 | M_{ik}=1, R_k=1$) represents the case when $M^*$ is correct, and $\underset{k\in \mathbf{K}_{-k'}}{\sum}Pr(M_{ik}^*=1, R_{k}=1 | M_{ik'}=1, R_{k'}=1)$ is the case where $M^*$ is misclassified. The sum over $k\in \mathbf{K}_{-k'}$ is the sum over all networks that are not the true network $k'$.
    \item Network selection, $M_{ik}$, is independent from intervention assignment, $R_k$; each network is randomized independently.
    \item Given that $P_M = Pr(M_{ik}^*=1|M_{ik}=1)$ is the probability of being classified into the correct network, $1-P_M=1-Pr(M_{ik}^*=1|M_{ik}=1)=Pr(M_{ik}^*=0|M_{ik}=1)$ is the probability of being misclassified into any network that is not one's true network, and $Pr(M_{ik}^*=1|M_{ik'}=1),\, k'\ne k$ is the probability of being misclassified into network $k$ specifically, given that the true network is $k'$, such that $\underset{k\in\mathbf{K}_{-k'}}{\sum}Pr(M_{ik}^*=1|M_{ik'}=1)=1-Pr(M_{ik'}^*=1|M_{ik'}=1)=1-P_M$.
\end{itemize}

\noindent Similarly, \vspace{-0.5cm}

\begin{flalign*}
\phi &= Pr(G_{ik}^*=0|G_{ik}=0) \\
& = Pr(M_{ik}^*=1, R_k=0 | M_{ik}=1, R_k=0) + \underset{k\in \mathbf{K}_{-k'}}{\sum}Pr(M_{ik}^*=1, R_{k}=0 | M_{ik'}=1, R_{k'}=0) \\
& = Pr(M_{ik}^*=1| M_{ik}=1, R_k=0) \\
&\quad +\underset{k\in \mathbf{K}_{-k'}}{\sum} Pr(M_{ik}^*=1| M_{ik'}=1, R_{k'}=0)Pr(R_{k}=0|M_{ik'}=1, R_{k'}=0) \\
& = Pr(M_{ik}^*=1|M_{ik}=1) + \underset{k\in \mathbf{K}_{-k'}}{\sum}Pr(M_{ik}^*=1|M_{ik'}=1)Pr(R_{k}=0) \\
& = P_M + (1-P_M)(1-P_R) &%
\end{flalign*}

\subsection*{Appendix 2: Derivations of probabilities and estimators in the main study}

The derivations assume non-differential misclassification ($Y_{ik}\perp G^*_{ik}|G_{ik}$). Recall the baseline outcome rate is defined as $Pr(Y_{ik}=1|G_{ik}=0)=P_{Y0}$, and $Pr(Y_{ik}=1|G_{ik}=1)=P_{Y0}\delta_{RR}$. Sensitivity is defined as $Pr(G^*_{ik}=1|G_{ik}=1)$, and specificity is $Pr(G^*_{ik}=0|G_{ik}=0)$. 

\begin{flalign*}
Pr(\mathpzc{A}) & = Pr(\mathpzc{A}|G_{ik}=1)Pr(G_{ik}=1)+Pr(\mathpzc{A}|G_{ik}=0)Pr(G_{ik}=0) \\
& = Pr(Y_{ik}=1,G^*_{ik}=1|G_{ik}=1)Pr(G_{ik}=1)+Pr(Y=1,G^*_{ik}=1|G_{ik}=0)Pr(G_{ik}=0)  \\
& = Pr(Y_{ik}=1|G_{ik}=1)Pr(G^*_{ik}=1|G_{ik}=1)Pr(G_{ik}=1) \\
&\quad + Pr(Y_{ik}=1|G_{ik}=0)Pr(G^*_{ik}=1|G_{ik}=0)Pr(G_{ik}=0) \\
& = (P_{Y0}\delta_{RR})\theta P_R + P_{Y0}(1-\phi)(1-P_R)\bigg\rvert_{\theta=P_M+(1-P_M)P_R,\; \phi=P_M+(1-P_M)(1-P_R)} \\
& = P_{Y0}P_R \{ \delta_{RR}[P_M+(1-P_M)P_R] + (1-P_M)(1-P_R)\} &%
\end{flalign*}

\begin{flalign*}
Pr(\mathpzc{B}) & = Pr(\mathpzc{B}|G_{ik}=1)Pr(G_{ik}=1)+Pr(\mathpzc{B}|G_{ik}=0)Pr(G_{ik}=0) \\
& = Pr(Y_{ik}=1,G^*_{ik}=0|G_{ik}=1)Pr(G_{ik}=1)+Pr(Y_{ik}=1,G^*_{ik}=0|G_{ik}=0)Pr(G_{ik}=0)  \\
& = Pr(Y_{ik}=1|G_{ik}=1)Pr(G^*_{ik}=0|G_{ik}=1)Pr(G_{ik}=1) \\
&\quad + Pr(Y_{ik}=1|G_{ik}=0)Pr(G^*_{ik}=0|G_{ik}=0)Pr(G_{ik}=0)\\
& = (P_{Y0}\delta_{RR})(1-\theta) P_R + P_{Y0}\phi(1-P_R)\bigg\rvert_{\theta=P_M+(1-P_M)P_R,\; \phi=P_M+(1-P_M)(1-P_R)} \\
& = P_{Y0} \{ P_R \delta_{RR}[1-(P_M + (1-P_M)P_R)] + [P_M + (1-P_M)(1-P_R)](1-P_R) \}  &%
\end{flalign*}

\begin{flalign*}
Pr(\mathpzc{N}_1) & = Pr(G^*_{ik}=1) \\
& = Pr(G^*_{ik}=1|G_{ik}=1)Pr(G_{ik}=1) + Pr(G^*_{ik}=1|G_{ik}=0)Pr(G_{ik}=0) \\
& = \theta P_R + (1-\phi)(1-P_R) \bigg\rvert_{\theta=P_M+(1-P_M)P_R,\; \phi=P_M+(1-P_M)(1-P_R)} \\
& = P_R &%
\end{flalign*}

\begin{flalign*}
Pr(\mathpzc{N}_0) & = Pr(G^*_{ik}=0) \\
& = Pr(G^*_{ik}=0|G_{ik}=1)Pr(G_{ik}=1) + Pr(G^*_{ik}=0|G_{ik}=0)Pr(G_{ik}=0) \\
& = (1-\theta) P_R + \phi(1-P_R)\bigg\rvert_{\theta=P_M+(1-P_M)P_R,\; \phi=P_M+(1-P_M)(1-P_R)} \\
& = 1-P_R &%
\end{flalign*} 

\begin{flalign*}
\delta_{RD}^* & = \frac{\mathbb{E}[\mathpzc{A}|N]}{\mathbb{E}[\mathpzc{N}_1|N]} - \frac{\mathbb{E}[\mathpzc{B}|N]}{\mathbb{E}[\mathpzc{N}_0|N]} \\
& = P_{Y0}\{ \delta_{RR}[P_M+(1-P_M)P_R] + (1-P_M)(1-P_R) \} \\  & \ \ \ - \frac{P_{Y0} \{ P_R \delta_{RR}[1-(P_M + (1-P_M)P_R)]+[P_M + (1-P_M)(1-P_R)](1-P_R)\}}{1-P_R} \\
& = \frac{P_{Y0}\delta_{RR}P_R[-P_M + (1-P_M)(1-P_R))-1+P_M+(1-P_M)P_R]}{1-P_R}\\
&\quad + \frac{P_{Y0}\delta_{RR}P_R[P_{Y0}\delta_{RR}P_M - P_{Y0}P_M(1-P_R)]}{1-P_R}\\
& = \frac{-P_{Y0}\delta_{RR}P_RP_M+P_{Y0}\delta_{RR}P_M-P_{Y0}P_M+P_{Y0}P_MP_R}{1-P_R} \\
& = \frac{P_{Y0}\delta_{RR}P_M(1-P_R)-P_{Y0}P_M(1-P_R)}{1-P_R} \\
& = P_{Y0}P_M(\delta_{RR}-1) &%
\end{flalign*}

\begin{flalign*}
\delta_{RR}^* & = \frac{\mathbb{E}[\mathpzc{A}|N]\mathbb{E}[\mathpzc{N}_0|N]}{\mathbb{E}[\mathpzc{B}|N]\mathbb{E}[\mathpzc{N}_1|N]}\\
& = \frac{\{\delta_{RR}[P_M+(1-P_M)P_R] + (1-P_M)(1-P_R)\}(1-P_R)}{ P_R \delta_{RR}[1-(P_M + (1-P_M)P_R)] + [P_M + (1-P_M)(1-P_R)](1-P_R)} &%
\end{flalign*} 

\vspace{0.5cm}

\noindent Asymptotically, by the law of large numbers, the empirical moments $\mathpzc{A}\xrightarrow[]{P}\mathbb{E}[\mathpzc{A}]$, $\mathpzc{B}\xrightarrow[]{P}\mathbb{E}[\mathpzc{B}]$, $N_1\xrightarrow[]{P}\mathbb{E}[\mathpzc{N}_1]$, and $N_0\xrightarrow[]{P}\mathbb{E}[\mathpzc{N}_0]$, such that $\widehat{\delta}_{RD}^* \xrightarrow[]{P}\delta^*_{RD}$ and $\widehat{\delta}_{RR}^* \xrightarrow[]{P}\delta^*_{RR}$ by Slutsky's theorem. 

\subsection*{Appendix 3: Derivation of bias}

\begin{flalign*}
\text{Bias}_{\widehat{\delta_{RD}^*}} & = \delta^*_{RD}-\delta_{RD} \\
& = P_{Y0}P_M(\delta_{RR}-1)- (P_{Y0} \times \delta_{RR} - P_{Y0}) \\
& = P_{Y0} (\delta_{RR}-1)(P_M-1) &%
\end{flalign*}

\begin{flalign*}
\text{Rel. Bias}_{\widehat{\delta_{RD}^*}} & = \displaystyle\left\lvert \frac{\delta^*_{RD}-\delta_{RD}}{\delta_{RD}} \right\rvert \\
& = \displaystyle\left\lvert \frac{P_{Y0}P_M(\delta_{RR}-1)- (P_{Y0} \times \delta_{RR} - P_{Y0})}{P_{Y0} \times \delta_{RR} - P_{Y0}} \right\rvert  \\
& = 1-P_M  &%
\end{flalign*}

\begin{flalign*}
\text{Bias}_{\widehat{\delta_{RR}^*}} & = \delta^*_{RR}-\delta_{RR} \\
& = \frac{\{\delta_{RR}[P_M+(1-P_M)P_R] + (1-P_M)(1-P_R)\}(1-P_R)}{ P_R \delta_{RR}[1-(P_M + (1-P_M)P_R)] + [P_M + (1-P_M)(1-P_R)](1-P_R)} - \delta_{RR} &%
\end{flalign*}

\begin{flalign*}
\text{Rel. Bias}_{\widehat{\delta_{RR}^*}} & = \displaystyle\left\lvert \frac{\delta^*_{RR}-\delta_{RR}}{\delta_{RR}} \right\rvert \\
& = \displaystyle\left\lvert \frac{\frac{\{\delta_{RR}[P_M+(1-P_M)P_R] + (1-P_M)(1-P_R)\}(1-P_R)}{ P_R \delta_{RR}[1-(P_M + (1-P_M)P_R)] + [P_M + (1-P_M)(1-P_R)](1-P_R)}- \delta_{RR}}{\delta_{RR}} \right\rvert \\
& = \displaystyle\left\lvert \frac{\{\delta_{RR}[P_M+(1-P_M)P_R] + (1-P_M)(1-P_R)\}(1-P_R)}{\delta_{RR}\{P_R \delta_{RR}[1-(P_M + (1-P_M)P_R)] + [P_M + (1-P_M)(1-P_R)](1-P_R)\}} -1 \right\rvert &%
\end{flalign*} 

\subsection*{Appendix 4: Derivation of bias-corrected estimators using the matrix method}

Recall that under $G$, $\delta_{RD}=\frac{\mathpzc{a}}{n_1}-\frac{\mathpzc{b}}{n_0}$, and $\delta_{RR}=\frac{\mathpzc{a}}{n_1}/\frac{\mathpzc{b}}{n_0}$ are the true ASpE (refer to Table~\ref{t:tableone}).
Under $G^*_{ik}$, $\widehat{\delta}_{RD}=\frac{\mathpzc{A}}{\mathpzc{N}_1}-\frac{\mathpzc{B}}{\mathpzc{N}_0}$, and $\widehat{\delta}_{RR}=\frac{\mathpzc{A}}{\mathpzc{N}_1}/\frac{\mathpzc{B}}{\mathpzc{N}_0}$, which are biased because $\mathpzc{A}$, $\mathpzc{B}$, $\mathpzc{N}_1$ and $\mathpzc{N}_0$ may all be mismeasured (refer to Table~\ref{t:tablethree}). The goal is to estimate $\widehat{\mathpzc{a}}$, $\widehat{\mathpzc{b}}$, $\widehat{n_1}$, and $\widehat{n_0}$ in order to estimate the corrected ASpE. The following is derived under Assumption 6, and assuming no misclassification of outcomes, such that $m_1$ is fixed. 

\begin{flalign*}
\mathbb{E}[\mathpzc{A}|m_1] & = Pr(G^*_{ik}=1|Y_{ik}=1)m_1 \\
& = Pr(G^*_{ik}=1,G_{ik}=1|Y_{ik}=1)m_1 + Pr(G^*_{ik}=1,G_{ik}=0|Y_{ik}=1)m_1 \\
& = Pr(G^*_{ik}=1|G_{ik}=1,Y_{ik}=1)Pr(G_{ik}=1|Y_{ik}=1)m_1 \\
&\quad + Pr(G^*_{ik}=1|G_{ik}=0,Y_{ik}=1)Pr(G_{ik}=0|Y_{ik}=1)m_1 \\
& = Pr(G^*_{ik}=1|G_{ik}=1)Pr(G_{ik}=1|Y_{ik}=1)m_1 + Pr(G^*_{ik}=1|G_{ik}=0)Pr(G_{ik}=0|Y_{ik}=1)m_1 \\
& = \theta \mathbb{E}[\mathpzc{a}|m_1] + (1-\phi)\mathbb{E}[\mathpzc{b}|m_1] &%
\end{flalign*}

\begin{flalign*}
\mathbb{E}[\mathpzc{B}|m_1] & = Pr(G^*_{ik}=0|Y_{ik}=1)m_1 \\
& = Pr(G^*_{ik}=0,G_{ik}=1|Y_{ik}=1)m_1 + Pr(G^*_{ik}=0,G_{ik}=0|Y_{ik}=1)m_1 \\
& = Pr(G^*_{ik}=0|G_{ik}=1,Y_{ik}=1)Pr(G_{ik}=1|Y_{ik}=1)m_1 \\
&\quad + Pr(G^*_{ik}=0|G_{ik}=0,Y_{ik}=1)Pr(G_{ik}=0|Y_{ik}=1)m_1 \\
& = Pr(G^*_{ik}=0|G_{ik}=1)Pr(G_{ik}=1|Y_{ik}=1)m_1 + Pr(G^*_{ik}=0|G_{ik}=0)Pr(G_{ik}=0|Y_{ik}=1)m_1 \\
& = (1-\theta)\mathbb{E}[\mathpzc{a}|m_1] + \phi \mathbb{E}[\mathpzc{b}|m_1] &%
\end{flalign*}

\noindent Solving the two equations with two unknowns, rewriting in matrix form:

\begin{flalign*}
\begin{bmatrix}
\mathbb{E}[\mathpzc{A}|m_1] \\
\mathbb{E}[\mathpzc{B}|m_1]
\end{bmatrix}
& =
\begin{bmatrix}
\theta & 1-\phi \\
1-\theta & \phi
\end{bmatrix} 
\begin{bmatrix}
\mathbb{E}[\mathpzc{a}|m_1] \\
\mathbb{E}[\mathpzc{b}|m_1]
\end{bmatrix} &%
\end{flalign*}

\begin{flalign*}
\begin{bmatrix}
\mathbb{E}[\mathpzc{a}|m_1] \\
\mathbb{E}[\mathpzc{b}|m_1]
\end{bmatrix}
& =
\begin{bmatrix}
\theta & 1-\phi \\
1-\theta & \phi
\end{bmatrix}^{-1}
\begin{bmatrix}
\mathbb{E}[\mathpzc{A}|m_1] \\
\mathbb{E}[\mathpzc{B}|m_1]
\end{bmatrix} \\
&=\frac{1}{\theta \phi - (1-\phi)(1-\theta)}
\begin{bmatrix}
\phi & -(1-\phi) \\
-(1-\theta) & \theta
\end{bmatrix}
\begin{bmatrix}
\mathbb{E}[\mathpzc{A}|m_1]\\
\mathbb{E}[\mathpzc{B}|m_1] 
\end{bmatrix}\\
&= \frac{1}{\theta \phi - (1-\theta-\phi + \theta \phi)}
\begin{bmatrix}
\phi & \phi-1 \\
\theta-1 & \theta
\end{bmatrix}
\begin{bmatrix}
\mathbb{E}[\mathpzc{A}|m_1]\\
\mathbb{E}[\mathpzc{B}|m_1] 
\end{bmatrix}\\
&= \frac{1}{\theta + \phi -1}
\begin{bmatrix}
\phi & \phi-1 \\
\theta-1 & \theta
\end{bmatrix}
\begin{bmatrix}
\mathbb{E}[\mathpzc{A}|m_1]\\
\mathbb{E}[\mathpzc{B}|m_1] 
\end{bmatrix} &%
\end{flalign*}

\begin{flalign*}
\mathbb{E}[\mathpzc{a}|m_1] & = \frac{\phi\mathbb{E}[\mathpzc{A}|m_1]+(\phi-1)\mathbb{E}[\mathpzc{B}|m_1]}{\theta+\phi-1}\overset{\mathpzc{A}+\mathpzc{B}=m_1}{=}\frac{\mathbb{E}[\mathpzc{B}|m_1]-\phi m_1}{1-\phi-\theta} \Rightarrow \mathpzc{\widehat{a}}=\frac{\mathpzc{B}-\widehat{\phi}m_1}{1-\widehat{\phi}-\widehat{\theta}} &%
\end{flalign*}

\begin{flalign*}
\mathbb{E}[\mathpzc{b}|m_1] &= \frac{(\theta-1)\mathbb{E}[\mathpzc{A}|m_1]+\theta\mathbb{E}[\mathpzc{B}|m_1]}{\theta+\phi-1}\overset{\mathpzc{A}+\mathpzc{B}=m_1}{=}\frac{\mathbb{E}[\mathpzc{A}|m_1]-\theta m_1}{1-\phi-\theta} \Rightarrow \mathpzc{\widehat{b}} =  \frac{\mathpzc{A}-\widehat{\theta}m_1}{1-\widehat{\phi}-\widehat{\theta}} &%
\end{flalign*}

\vspace{0.5cm}

\noindent Similarly, $\mathbb{E}[c|m_0]=\frac{\mathbb{E}[D|m_0]-\phi m_0}{1-\phi-\theta}$ and $\mathbb{E}[d|m_0]=\frac{\mathbb{E}[C|m_0]-\theta m_0}{1-\phi-\theta}$. Then,

\begin{flalign*}
\mathbb{E}[\mathpzc{N}_1|N] &= \mathbb{E}[\mathpzc{a}|m_1]+ \mathbb{E}[\mathpzc{c}|m_0]=\frac{\mathbb{E}[\mathpzc{B}|m_1]+\mathbb{E}[\mathpzc{D}|m_0]-\phi \mathbb{E}[m_1+m_0]}{1-\phi-\theta}=\frac{\mathbb{E}[N_0|N]-\phi N}{1-\phi-\theta}&%
\end{flalign*}

\begin{flalign*}
\mathbb{E}[N_0|N] &= \mathbb{E}[\mathpzc{b}|m_1]+ \mathbb{E}[\mathpzc{d}|m_0]=\frac{\mathbb{E}[\mathpzc{A}|m_1]+\mathbb{E}[\mathpzc{C}|m_0]-\theta \mathbb{E}[m_1+m_0]}{1-\phi-\theta}=\frac{\mathbb{E}[\mathpzc{N}_1|N]-\theta N}{1-\phi-\theta}&%
\end{flalign*}

\begin{flalign*}
\widehat{n_1}&=\frac{N_0-\widehat{\phi}N}{1-\widehat{\phi}-\widehat{\theta}}\quad ; \quad \widehat{n_0}=\frac{N_1-\widehat{\theta}N}{1-\widehat{\phi}-\widehat{\theta}} &%
\end{flalign*}

\vspace{0.5cm}

\noindent By substituting the observed with the corrected cell counts, we obtain the corrected ASpE estimates. Since $\mathpzc{A}$, $\mathpzc{B}$, $\mathpzc{N}_1$, and $\mathpzc{N}_0$ are empirical moments as described in Appendix 2, as are $\widehat{\theta}$ and $\widehat{\phi}$ given they are also estimated from a $2\times2$ table, the estimators of the bias-corrected ASpE are consistent, $\widehat{\delta}_{RD} \xrightarrow[]{P} \delta_{RD}$ and $\widehat{\delta}_{RR} \xrightarrow[]{P} \delta_{RR}$. 

\subsection*{Appendix 5: Restrictions of the matrix method}

Recall bias-corrected cell counts using the matrix method are the following:\\ $\widehat{a}=\frac{\mathpzc{B}-\widehat{\phi}m_1}{1-\widehat{\phi}-\widehat{\theta}}$, $\widehat{b}=\frac{\mathpzc{A}-\widehat{\theta}m_1}{1-\widehat{\phi}-\widehat{\theta}}$, $\widehat{c}=\frac{\mathpzc{D}-\widehat{\phi}m_0}{1-\widehat{\phi}-\widehat{\theta}}$, $\widehat{d}=\frac{\mathpzc{C}-\widehat{\theta}m_1}{1-\widehat{\phi}-\widehat{\theta}}$\\

\noindent For $\widehat{\theta}+\widehat{\phi}>1$, the denominator is negative. To obtain positive cell counts, the numerator also needs to be negative, i.e. $\mathpzc{A}-\widehat{\theta}m_1 <0 \Rightarrow \widehat{\theta}>\frac{\mathpzc{A}}{m_1}$, and $\mathpzc{C}-\widehat{\theta}m_0 <0 \Rightarrow \widehat{\theta}>\frac{\mathpzc{C}}{m_0}$. Therefore, $\widehat{\theta}>\text{max}(\frac{\mathpzc{A}}{m_1},\frac{\mathpzc{C}}{m_0})$.\\

\noindent Similarly, $\mathpzc{B}-\widehat{\phi}m_1 <0 \Rightarrow \widehat{\phi}>\frac{\mathpzc{B}}{m_1}$, and $\mathpzc{D}-\widehat{\phi}m_0 <0 \Rightarrow \widehat{\phi}>\frac{\mathpzc{D}}{m_0}$. Therefore, $\widehat{\phi}>\text{max}(\frac{\mathpzc{B}}{m_1},\frac{\mathpzc{D}}{m_0})$.\\

\noindent The same logic follows for $\widehat{\theta}+\widehat{\phi}<1$. In this case, the denominator is positive. To obtain positive cell counts, the numerator also needs to be positive, i.e. $\mathpzc{A}-\widehat{\theta}m_1 >0 \Rightarrow \widehat{\theta}<\frac{\mathpzc{A}}{m_1}$, and $\mathpzc{C}-\widehat{\theta}m_0 >0 \Rightarrow \widehat{\theta}<\frac{\mathpzc{C}}{m_0}$. Therefore, $\widehat{\theta}<\text{min}(\frac{\mathpzc{A}}{m_1},\frac{\mathpzc{C}}{m_0})$.\\

\noindent Likewise, $\mathpzc{B}-\widehat{\phi}m_1 >0 \Rightarrow \widehat{\phi}<\frac{\mathpzc{B}}{m_1}$, and $\mathpzc{D}-\widehat{\phi}m_0 >0 \Rightarrow \widehat{\phi}<\frac{\mathpzc{D}}{m_0}$. Therefore, $\widehat{\phi}<\text{min}(\frac{\mathpzc{B}}{m_1},\frac{\mathpzc{D}}{m_0})$.

\subsection*{Appendix 6: Derivation of bias-corrected estimators using the inverse matrix method}

Recall $\theta=Pr(G^*_{ik}=1|G_{ik}=1)$ and $\phi=Pr(G^*_{ik}=0|G_{ik}=0)$, the folloinwg is derived under Assumption 6.

\begin{flalign*}
\mathbb{E}[\mathpzc{a}|m_1] & = Pr(G_{ik}=1|Y_{ik}=1)m_1 \\
& = Pr(G_{ik}=1,G^*_{ik}=1|Y_{ik}=1)m_1 + Pr(G_{ik}=1,G^*_{ik}=0|Y_{ik}=1)m_1 \\
& = Pr(G_{ik}=1|G^*_{ik}=1,Y_{ik}=1)Pr(G^*_{ik}=1|Y_{ik}=1)m_1 \\
&\quad + Pr(G_{ik}=1|G^*_{ik}=0,Y_{ik}=1)Pr(G^*_{ik}=0|Y_{ik}=1)m_1 \\
& = PPV_1 \mathbb{E}[\mathpzc{A}|m_1] + (1-NPV_1)\mathbb{E}[\mathpzc{B}|m_1] &%
\end{flalign*}

Then, $\widehat{a}=\widehat{\mathbb{E}}[a|m_1]=\widehat{PPV}_1\mathpzc{A}+(1-\widehat{NPV}_1)\mathpzc{B}$. Similarly, $\widehat{b}=(1-\widehat{PPV}_1)\mathpzc{A}+\widehat{NPV}_1\mathpzc{B}$, $\widehat{c}=\widehat{PPV}_0\mathpzc{C}+(1-\widehat{NPV}_0)\mathpzc{D}$, $\widehat{d}=(1-\widehat{PPV}_0)\mathpzc{C}+\widehat{NPV}_0\mathpzc{D}$.

\vspace{0.5cm}
\noindent Using the estimated cell counts, the bias-corrected ASpE risk difference and risk ratio can be obtained.

\subsection*{Appendix Tables}

\begin{table}[h!]
\centering
\caption*{Appendix Table 1: Effect of misclassification probability ($1-P_M$) on the bias of $\widehat{\delta}^*_{RD}$ and $\widehat{\delta}^*_{RR}$ when outcomes are not correlated, given as empirical averages over $P_{Y0} \in (0.1, 0.25, 0.5, 0.75, 0.9), \delta_{RR} \in (0.25, 0.75, 1.25, 3, 5), P_R \in (0.2, 0.5, 0.8)$, with empirical standard errors (SE)} 
{\begin{tabular*}{\textwidth}{@{}l@{\extracolsep{\fill}}c@{\extracolsep{\fill}}c@{\extracolsep{\fill}}c@{\extracolsep{\fill}}c@{\extracolsep{\fill}}c@{}}
\hline 
$1-P_M$ & $\text{Bias}_{\widehat{\delta}^*_{RD}}$ (SE) & $\text{Relative Bias}_{\widehat{\delta}^*_{RD}}$ (SE) & $\text{Bias}_{\widehat{\delta}^*_{RR}}$ (SE) & $\text{Relative Bias}_{\widehat{\delta}^*_{RR}}$ (SE) \\
    \hline
    0.10 & 0.01 (0.01) & 0.10 (0.05) & 0.02 (0.03) & 0.08 (0.03)\\ 
    0.25 & 0.02 (0.02) & 0.25 (0.08) & 0.06 (0.05) & 0.12 (0.06)\\  
    0.50 & 0.03 (0.02) & 0.50 (0.10) & 0.11 (0.06) & 0.20 (0.07)\\ 
    0.75 & 0.05 (0.03) & 0.75 (0.11) & 0.18 (0.08) & 0.32 (0.09)\\ 
    0.90 & 0.06 (0.03) & 0.90 (0.12) & 0.22 (0.10) & 0.37 (0.11)\\ 
    \hline 
 \end{tabular*}
\vspace*{-6pt}}
\end{table}

\begin{table}[h!]
\centering
\caption*{Appendix Table 2: Effect of baseline outcome rate ($P_{Y0}$) on the bias of $\widehat{\delta}^*_{RD}$ and $\widehat{\delta}^*_{RR}$ when outcomes are not correlated, given as empirical averages over $P_M \in (0.1, 0.25, 0.5, 0.75, 0.9), \delta_{RR} \in (0.25, 0.75, 1.25, 3, 5), P_R \in (0.2, 0.5, 0.8)$, with empirical standard errors (SE)} 
{\begin{tabular*}{\textwidth}{@{}l@{\extracolsep{\fill}}c@{\extracolsep{\fill}}c@{\extracolsep{\fill}}c@{\extracolsep{\fill}}c@{\extracolsep{\fill}}c@{}}
\hline 
$P_{y0}$ & $\text{Bias}_{\widehat{\delta}^*_{RD}}$ (SE) & $\text{Relative Bias}_{\widehat{\delta}^*_{RD}}$ (SE) & $\text{Bias}_{\widehat{\delta}^*_{RR}}$ (SE) & $\text{Relative Bias}_{\widehat{\delta}^*_{RR}}$ (SE)  \\
    \hline
    0.10 & -0.01 (0.02) & 0.75 (0.12) & -0.14 (0.20) & 0.34 (0.08)\\  
    0.25 & 0.00 (0.02) & 0.52 (0.12) & 0.00 (0.10) & 0.23 (0.07)\\ 
    0.50 & 0.04 (0.02) & 0.50 (0.14) & 0.06 (0.05) & 0.18 (0.06)\\ 
    0.75 & 0.06 (0.02) & 0.50 (0.07) & 0.06 (0.03) & 0.18 (0.04)\\ 
    0.90 & 0.17 (0.02) & 0.50 (0.04) & 0.17 (0.02) & 0.28 (0.05)\\ 
    \hline 
 \end{tabular*}
\vspace*{-6pt}}
\end{table}

\begin{table}[h!]
\centering
\caption*{Appendix Table 3: Effect of spillover effect ($\delta_{RR}$) on the bias of $\widehat{\delta}^*_{RD}$ and $\widehat{\delta}^*_{RR}$ when outcomes are not correlated, given as empirical averages over $P_M \in (0.1, 0.25, 0.5, 0.75, 0.9), P_{Y0} \in (0.1, 0.25, 0.5, 0.75, 0.9), P_R \in (0.2, 0.5, 0.8)$, with empirical standard errors (SE)} 
{\begin{tabular*}{\textwidth}{@{}l@{\extracolsep{\fill}}c@{\extracolsep{\fill}}c@{\extracolsep{\fill}}c@{\extracolsep{\fill}}c@{\extracolsep{\fill}}c@{}}
\hline 
$\delta_{RR}$ & $\text{Bias}_{\widehat{\delta}^*_{RD}}$ (SE) & $\text{Relative Bias}_{\widehat{\delta}^*_{RD}}$ (SE) & $\text{Bias}_{\widehat{\delta}^*_{RR}}$ (SE) & $\text{Relative  Bias}_{\widehat{\delta}^*_{RR}}$ (SE)  \\
    \hline
    0.25 & 0.14 (0.02) & 0.50 (0.05) & 0.29 (0.04) & 1.16 (0.23)\\  
    0.75 & 0.05 (0.02) & 0.58 (0.15) & 0.12 (0.04) & 0.16 (0.06)\\ 
    1.25 & -0.03 (0.02) & 0.75 (0.24) & -0.13 (0.06) & 0.11 (0.04)\\ 
    3.00 & -0.14 (0.02) & 0.50 (0.05) & -1.35 (0.22) & 0.44 (0.04)\\ 
    5.00 & -0.20 (0.02) & 0.50 (0.05) & -3.04 (0.49) & 0.60 (0.04)\\ 
    \hline 
 \end{tabular*}
\vspace*{-6pt}}
\end{table}

\begin{table}[h!]
\centering
\caption*{Appendix Table 4: Effect of intervention allocation probability ($P_R$) on the bias of $\widehat{\delta}^*_{RD}$ and $\widehat{\delta}^*_{RR}$ when outcomes are not correlated, given as empirical averages over $P_M \in (0.1, 0.25, 0.5, 0.75, 0.9), P_{Y0} \in (0.1, 0.25, 0.5, 0.75, 0.9), \delta_{RR} \in (0.25, 0.75, 1.25, 3, 5)$, with empirical standard errors (SE)} 
{\begin{tabular*}{\textwidth}{@{}l@{\extracolsep{\fill}}c@{\extracolsep{\fill}}c@{\extracolsep{\fill}}c@{\extracolsep{\fill}}c@{\extracolsep{\fill}}c@{}}
\hline 
$P_R$ & $\text{Bias}_{\widehat{\delta}^*_{RD}}$ (SE) & $\text{Relative Bias}_{\widehat{\delta}^*_{RD}}$ (SE) & $\text{Bias}_{\widehat{\delta}^*_{RR}}$ (SE) & $\text{Relative Bias}_{\widehat{\delta}^*_{RR}}$ (SE)  \\
    \hline
    0.20 & 0.02 (0.02) & 0.50 (0.10) & 0.06 (0.07) & 0.26 (0.07)\\  
    0.50 & 0.02 (0.02) & 0.50 (0.08) & 0.05 (0.05) & 0.24 (0.06)\\ 
    0.80 & 0.02 (0.02) & 0.51 (0.10) & 0.03 (0.06) & 0.23 (0.07)\\ 
    \hline 
 \end{tabular*}
\vspace*{-6pt}}
\end{table}

\end{document}